\documentclass[journal ]{new-aiaa}
\usepackage[utf8]{inputenc}
\usepackage{textcomp}
\usepackage{multicol, blindtext}
\usepackage{subfig}
\usepackage[export]{adjustbox}
\usepackage{graphicx}
\usepackage{color}
\usepackage{cancel}

\usepackage{soul}

\usepackage{comment}
\usepackage{floatrow}
\usepackage{float}
\floatstyle{plaintop}
\restylefloat{table}
\usepackage{amsmath}
\usepackage{mathtools}
\usepackage[version=4]{mhchem}
\usepackage{siunitx}
\usepackage{longtable,tabularx}
\usepackage[export]{adjustbox}
\usepackage[ruled,vlined]{algorithm2e}

\usepackage{amsmath}
\usepackage{booktabs}
\usepackage{varwidth}

\newtheorem{remark}{Remark}[section]

\newtheorem{problem}{Problem}[section]

\title{Station-keeping of $L_2$ halo orbits under sampled-data model predictive control}

\author{Mohamed Elobaid\footnote{PhD student, DIAG and L2S, mohamed.elobaid@uniroma1.it, mohamed.elobaid@centralesupelec.fr, corresponding author.},  Mattia Mattioni\footnote{Researcher, DIAG, mattia.mattioni@uniroma1.it.} and  Salvatore Monaco\footnote{Full Professor, DIAG,  salvatore.monaco@uniroma1.it.}}
\affil{Department of Computer, Control and Management Engineering (DIAG) (La Sapienza university of Rome); via Ariosto 25, 00185 Rome (Italy)}
\author{Doroth\'ee Normand-Cyrot\footnote{Directeur de Recherche, L2S (CNRS), dorothee.normand-cyrot@centralesupelec.fr.}}
\affil{Laboratory of Signals and Systems (L2S, UMR 8506) (CNRS and University of Paris Saclay); 3, Rue Joliot Curie, 91192, Gif-sur-Yvette (France)}

\begin{document}

\maketitle

\section{Introduction}


The dynamics of a spacecraft in the Earth-Moon gravitational system (the so-called Restricted Three-Body Problem (RTBP) \cite{Poincare, Barrow}) exhibit equilibria commonly known as Libration (or Lagrangian) points. This makes Lagrangian points excellent locations for spacecrafts in exploration applications \cite{NasaReport}. Among those the so-called \textit{translunar} Libration point $L_2$ has been attracting particular interest for various satellite communication applications \cite{comm_design} and deep space observation and exploration purposes \cite{NasaReport3} as also testified by the numerous space missions that have already taken place (e.g., NASA's ARTEMIS \cite{ARTEMIS} and CNSA's Chang'e 5-T1 \cite{Change5T1}) and new ones are in schedule \cite{NasaInterest}.

However, keeping a spacecraft close to $L_2$ requires an active control action because, as well-known, it is an unstable equilibrium point as are most equilibrium trajectories within the Halo family of orbits around it  \cite{GomezBook, zimovan}.  In this respect, several control design methods have been proposed throughout the last decades for stabilizing Halo orbits while mitigating the effect of perturbations and unmodelled dynamics (see \cite{survey} for a complete survey, and \cite{howell} for an earlier work). Among these and related to this work, the most important ones are based on optimization  \cite{Simo, Folta, Ratner, ARTEMIS2,  long-term, LQR3, bando}, receding horizon \cite{linearMPC, polynMPC} and advanced nonlinear design (e.g. projection to the stable manifold \cite {projection}, feedback linearization, nonlinear regulation \cite{Paolo} and backstepping \cite{Backstepping}). In the majority of the aforementioned works, a circular RTBP model is used to describe the dynamics of a spacecraft so neglecting the primaries motion perturbations whose effect, however, cannot be discarded in practice. Only few works consider the more realistic elliptic RTBPs in which the eccentricity is not zero (e.g. \cite{NasaReport,Paolo}) possibly making use of ephemeris models considering the solar radiation pressure and other  gravitational perturbations (e.g. \cite{Backstepping}). 

In the sequel, we provide a new control scheme assuming the so-called elliptic RTBP model under the influence of solar radiation pressure which captures the principal feature of the dynamics while retaining simplicity \cite{survey}. In addition, a modified model predictive control approach is proposed to cope with real-time implementation issues as, for instance, the digital nature of both actuation and sensing and possible limits on control, unavoidable in practice. In particular, we consider the case in which measures are available only at sporadic time instants whereas the control is piecewise constant over the sampling period (e.g. \cite{lorenzo}).  The approach we propose is based on the design of multi-rate (MR) strategies at the planning level providing a suitable reference governor for a model-predictive control (MPC) control scheme. The resulting control system enables to overcome the limits of small prediction horizons underlined in \cite{polynMPC}, when the controller is designed under usual MPC. As a matter of fact, such limits would bring to unfeasible controllers for small prediction horizons when the design makes use of a the sampled-data model, due to the cancellation of the possibly unstable zero-dynamics under sampling \cite{nolcos}. Also, the use of large prediction horizons, typically used as a stratagem in practice, might introduce large computational delays.

This work is contextualized in this framework by proposing a new sampled-data MPC control scheme where the problem is simplified. This is achieved by generating a suitable reference trajectory based on a multi-rate planner which guarantees stability of the closed loop and feasibility of the optimization problem solved, at each step, by the MPC assuming cheap control. In particular, such a planner is designed starting from the nonlinear regulation-based controller proposed in \cite{Paolo} which yields, by construction, admissible and bounded trajectories which can be then fed to the MPC as reference to track.
In this context, the contribution of this paper stands in the proposition of a new control scheme which employs at the low level an inherently robust nonlinear MPC fed by a references generated by a multi-rate sampled-data model of a control system designed making use of nonlinear regulation.

The choice of the proposed control scheme is motivated by the observation that MPC and nonlinear regulation can be employed together to mitigate each the deficiencies of the other: nonlinear regulation appears to be the natural context for setting the problem since the references and perturbations are periodic although it lacks in robustness to unmodelled disturbances; MPC is becoming a standard tool for handling tracking applications, it is inherently robust to bounded perturbations \cite{grune} despite it requiring reference signals pre-processing to guarantee convergence and recursive feasibility.
Thus, the implementation of a nonlinear MPC fed by samples of a reference signal generated by the sampled-data model of the controlled circular RTBP under nonlinear regulation provides a natural solution.

The remainder of this work is organized as follows. Notations are  given at the end of this section. Section \ref{SEC:II}  describes the motion equations, the class of quasi Halo orbit references under investigation and formally states the problem. Some background material on the employed control strategies which are at the basis of the proposed control scheme are detailed in Section \ref{SEC:III}. The proposed control scheme is developed in Section \ref{SEC:IV}. Section \ref{SEC:V} presents a comparative study highlighting the advantages of the proposed approach as well as brief discussion of computational aspects. Concluding remarks end this note. 

\subsection*{Notations}

    $\mathbb R$, $\mathbb{C}$ and $\mathbb{N}$ denote the sets of real, complex and natural numbers respectively whereas $\mathbb{Z}_{\geq 0} = \mathbb{N} \cup \{ 0\}$,  $\mathbb{S}^1 = \{ p \in \mathbb{C}: |p| < 1  \}$  and $\mathbb{C}^-$ is the left-hand side of the complex plane. For a real matrix $A\in \mathbb{R}^{n\times n}$, $\sigma(A)\subset \mathbb{C}$ denotes its spectrum. The symbols $>0$ and $<0$ (resp. $\succ$ and $\prec$) denote  positive and negative definite functions (resp. matrices), respectively. Given a vector $q \in \mathbb{R}^n$, and matrix $A\succeq 0$, the weighted norm of $q$ is denoted $\|q\|_A = \sqrt{q^\top A q}$. $I$ and $\text{Id}$ are the identity matrix (of suitable dimensions) and  the identity operator respectively.  All vector fields and mappings are assumed smooth. Given a smooth vector field $f$ over $\mathbb R^n$ the Lie derivative is denoted by $\mathrm{L}_f = \sum_{i= 1}^n f_i(\cdot)\frac{\partial}{\partial q_i}$ and recursively $\mathrm{L}_f^i = \mathrm{L}_f (\mathrm{L}_f^{i-1})$ with $\mathrm{L}_f^0 = \text{Id}$.  The Lie exponent operator is given by $e^{\mathrm{L}_f}:= \text{Id} + \sum_{i \geq 1}\frac{\mathrm{L}_f^i}{i!}$. A continuous function $\gamma: [0, \infty) \to [0, \infty)$ with $\gamma(0) = 0$ is said to be of class $\mathcal{K}_\infty$ if it is unbounded and strictly increasing. A continuous function $R(x,\delta)$ is of order $\mathcal{O}(\delta^p)$ with $p\geq1$ if, whenever it is defined, it can be written as $R(x,\delta) = \delta^{p-1}\tilde{R}(x,\delta)$ and there exists a function $\beta(\delta) \in K_{\infty}$ and $\delta^{\star} > 0$ such that $\forall \delta \leq \delta^{\star},  |\tilde{R}(x,\delta)| \leq \beta(\delta)$.  Both $0_{m \times n}, \ I_{n}$ denote a zero matrix of dimension $m \times n$ and the Identity matrix of dimension $n$ respectively.  


\section{Modelling and Problem Statement}\label{SEC:II}

\subsection{The spacecraft model}

The motion of a small third body under the gravitational pull of both the Earth and Moon gets the form (see \cite{survey} and the references therein) 
{\small{\begin{align}\nonumber
        \ddot x  \textcolor{black}{-u_1} &- 2\dot y(1+\beta(t)) - x(1+2\beta(t)+\beta(t)^2)-y\dot{\beta}(t) = -\frac{(1-\mu)(x+\mu(1+\alpha(t)))}{r_{1}}-\frac{\mu(x-(1-\mu)(1+\alpha(t)))}{r_{2}} + \frac{g_{sc}}{c} \cos^2(\zeta t) \\
        \ddot y   \textcolor{black}{-u_2} &- 2\dot x(1+\beta(t)) - y (1+2\beta(t)+\beta(t)^2) - x\dot \beta(t) = -\frac{(1-\mu)y}{r_{1}}-\frac{\mu y}{r_{2}} + \frac{g_{sc}}{c} \sin^2(\zeta t)
        \label{simulationmodel}
        \\
        \ddot z   \textcolor{black}{-u_3}&= -\frac{(1-\mu)z}{r_{1}}-\frac{\mu z}{r_{2}}\nonumber
\end{align}
}}
where: $e \approx 0.0549$ is the eccentricity; $x\, , y \, ,z$ are the  distance normalized coordinates of the third body expressed in a rotating frame centered at the barycenter when the eccentricity $e$ is  small; $u_1\, , u_2\, , u_3$ are three axis thrusts in the corresponding direction; $\alpha(t)$ and $\beta(t)$ are periodic functions with zero mean value admitting the form of a series expansion in $e$ given by
\begin{subequations}
    \label{eq:alpha_beta}
    \begin{align}
        \alpha(t) &= -e\cos(t+\phi) + 0.5e^2(1 + \cos(2t + 2\phi)) + O(e^3)\\
        \beta(t) &= 2e\cos(t+\phi) + 2.5e^2\cos(2t +2 \phi) +  O(e^3) ;
    \end{align}
\end{subequations}
$m_1, m_2$ are the masses of the Earth and Moon respectively, and $\mu = \frac{m_2}{m_1+m_2}$; $r_i = \frac{|r_{i,3}|}{\|r_{i,3}\|^3}$ for $ i = 1,2$ denote the norm, in the distance normalized rotating frame, of the vector from the Earth to the third body and from the moon to the third body respectively; $g_{sc}$ and $c$ are the solar constant and the speed of light with $\frac{g_{sc}}{c} \approx 4.5 \time 10^{-6} N/m^2$; finally, $\zeta \approx 0.9252$ is the angular rate of the sun light line in non-dimensional units.
\begin{remark}
It is assumed that the average distance between Earth and Moon is normalized, as well as the sum of masses of Earth and Moon (i.e., $m_1 + m_2$) and  the average angular rate of change between the rotating and inertial frame \cite{NasaReport}.
\end{remark}
\begin{remark}
The periodic functions $\alpha(t)$ and $\beta(t)$ are used to capture the effect of eccentricity $e$ of the orbits; as a matter of fact, when $\alpha(t) = \beta(t) = 0$,  (\ref{simulationmodel}) reduces to the well known approximated circular RTBP model. 
\end{remark}

The motion equations (\ref{simulationmodel}) can be re-written in the control affine perturbed state-space form 
\begin{equation}\label{ssmodel}
\begin{aligned}
    \dot q &= f(q, \xi) + {B(u+D)}\\
    p &= h(q) = Cq 
\end{aligned}
\end{equation}
with
 \begin{align*}
     f(q,\xi) =&  \begin{bmatrix}
     f_1(q) \\ f_2(q,\xi) 
     \end{bmatrix},\quad  f_1(q) = \begin{bmatrix}
     0_{3 \times 3} & I_3
     \end{bmatrix} q, \quad B = \begin{bmatrix}
     0_{3 \times 3} \\
     I_3
     \end{bmatrix}, \quad C = \begin{bmatrix} I_3 & 0_{3 \times 3} \end{bmatrix}
     \\  f_2(q, \xi) =& - M p - 2N\dot p - \xi_1(2Mp+2N\dot p) - \xi_2Mp - \xi_3 N\dot p  -  \frac{1-\mu}{\|p - d_1\|^3}(p - d_1) -  \frac{\mu}{\|p - d_2\|^3}(p - d_2)\\
    M =& \begin{bmatrix} -1 & 0 & 0\\0 & -1 & 0\\ 0& 0 & 0 \end{bmatrix}, \quad N = \begin{bmatrix} 0 & -1 & 0\\1 & 0 & 0\\ 0& 0 & 0 \end{bmatrix}, \quad d_1 = \begin{bmatrix} -\mu - \frac{\xi_4}{1-\mu}\\0 \\ 0\end{bmatrix}, \quad d_2 = \begin{bmatrix} 1-\mu + \frac{\xi_4}{\mu}\\0 \\ 0\end{bmatrix}
 \end{align*}
where $p =  (x \ y \ z)^T\in \mathbb{R}^3$ specifies the output,   $q := (q_1\  \hdots \  q_6)^{\top} =  (p^\top \ \dot p^\top )^\top \in \mathbb{R}^6$ the state, $u \in \mathbb{R}^3$ the control input, 
\begin{align}\label{eq:xi}
    \xi(t) = 
    \begin{pmatrix}\beta(t) & \beta^2(t) & \dot{\beta}(t) & \mu(1-\mu)\alpha(t)\end{pmatrix}^{\top} \in \mathbb{R}^4
\end{align}
 and the vector of the eccentricity perturbation and the Solar Radiation Pressure (SRP)
\begin{align*}
    D(t) = \begin{bmatrix} 
    \frac{G_{sc}}{c}\cos^2(\zeta t) \\ \frac{G_{sc}}{c}\sin^2(\zeta t) \\ 0 \end{bmatrix}.
\end{align*}


When setting $\xi = 0_{4 \times 1}$, $D(t) = 0_{3 \times 1}$, the dynamics \eqref{simulationmodel} reduces to the so-called circular RTBP model. In that case, the resultin dynamics possesses five equilibria with fixed location on the rotating frame and usually denoted by $L_i$, with $i = 1, \hdots, 5$.
Among these, $L_1$, $L_2$ and $L_3$ are the so-called collinear points with the corresponding $x$ coordinates, in the distance normalized rotating frame, given by $L_3 = -1.0050627, L_1 = 0.8369147$, $L_2 = 1.155682$. 
As shown in  \cite{Richardson, Orbits} (see also \cite{initialConditions, ae} for comments on the circular and elliptic RTBP cases respectively), the solutions of (\ref{ssmodel}) close to the collinear equilibria,  when suitably initialized, are periodic in nature as also underlined by the corresponding linear tangent model of (\ref{simulationmodel}). 

\subsection{The quasi-halo orbit model and $L_2$-station keeping}

When neglecting external signals (i.e., $D = u =  0_{3 \times 1}$), the linear tangent model of \eqref{ssmodel} at $q^\star = ( L_2 \ 0 \ 0 \ 0 \ 0 \ 0)^\top$ gets the form
\begin{equation}\label{refGen}
\begin{aligned}
    \dot q &= Aq , \quad  A = \begin{bmatrix} 0_{3 \times 3} & I_3 \\ A_{2,1} & A_{2,2} \end{bmatrix}, \quad A_{2,1} = \begin{bmatrix} 1+2\eta & 0 & 0  \\ 0 & 1-\eta & 0\\ 0 & 0 & -\eta \end{bmatrix}  , \quad A_{2,2} = \begin{bmatrix} 0 & 2 & 0 \\ -2 & 0 & 0\\ 0 & 0 & 0 \end{bmatrix}  
\end{aligned}
\end{equation}
with $\eta = \frac{1-\mu}{(L_2 + \mu)^3} + \frac{\mu}{(L_2 - 1 + \mu)^3} = 3.194075$. Because $A$ possesses periodic modes, for a suitable set of initial conditions $q_0 \in \mathbb{R}^6$,  (\ref{refGen}) admits periodic and bounded solutions of the form $q(t) = ( \nu(t) \ \dot \nu(t))^\top$  \cite{Richardson}  with
\begin{equation}\label{ref}
\begin{aligned}
\nu(t) = \begin{bmatrix} -\frac{k(1-\eta + \Omega^2)}{2\Omega} \cos (\Omega t + \phi) \quad
& k \sin (\Omega t + \phi) \quad
& k \cos (\Omega_z t) \end{bmatrix}^\top.
\end{aligned}
\end{equation}
where, in addition, $k$ being a constant depending on the initial displacement from $L_2$, $\phi$ a phase angle and $\Omega, \Omega_z$ the in-plane and out-of-plane natural frequencies. It is assumed that  $\Omega = \Omega_z = 1.8636$ for guaranteeing visibility from Earth.  In addition, when simulating, $\phi$ is set to zero. 

Expression (\ref{ref}) defines a \textit{quasi} Halo orbit that is an approximate circular orbit in 3D centered around $L_2$. In this sense, the objective is to drive and maintain the satellite onto such an orbit. Formally, the control objective consists in designing a feedback law ensuring tracking of the quasi halo orbit described by \eqref{ref} (i.e., that $p(t)\to \nu(t)$ as $t \to \infty$). In the concerned literature (e.g., \cite{survey} and references therein) such a problem is commonly referred to as \emph{station-keeping} of the $L_2$ orbit and has found, as already mentioned, numerous solutions under different working assumptions. However none of them considers, at the same time, issues arising from both digital implementation of the control laws and the effect of unmodelled perturbations over the closed loop as formally set here below.



\begin{problem}[$L_2$ station-keeping under digital control]\label{problem}
 Consider a spacecraft whose dynamics around $L_2$ is described by (\ref{ssmodel}). Assume that measures of the states are available at periodic sampling instants $t = k\delta$ with $k \in \mathbb{Z}_{\geq 0}$, and $\delta \geq 0$ the sampling period. In addition, let the the control be piecewise constant over intervals of length $\delta$; i.e., $u(t) = u(k\delta) = u(k)$ for $t\in [k\delta, (k+1)\delta ($.  
 Design a piecewise constant feedback $u(k) = \gamma(q(k), \nu(k))$ ensuring that the position of the satellite approaches and stays on the quasi-Halo orbit reference (\ref{ref}) at all sampling instants $t = k\delta , \hspace{1mm} k \geq 0$; namely, $ p(k\delta) \to \nu(k\delta)$ as  $k \to \infty$ and $p((k+1)\delta) = \nu((k+1)\delta)$ if $p(k\delta) = \nu(k\delta)$.
\end{problem}
Before going in the details of the solution we propose, the continuous-time one based on regulation is recalled from \cite{Paolo} together with preliminaries on sampling and nonlinear MPC.

\section{Recalls on nonlinear continuous-time and sampled-data $L_2$ station keeping control}\label{SEC:III}

\subsection{Continuous-time nonlinear regulation for Halo orbits stabilization}

When neglecting the effects of the SRP, a good solution to the $L_2$-station-keeping control is achieved under regulation by fixing the exosystem as
\begin{subequations}\label{exosystem}
\begin{align}
    \dot \omega =& S\omega\\
    \xi =& T_\xi \omega \label{eq:xi}\\
     \nu =& T_\nu \omega  \label{eq:nu}
\end{align}
\end{subequations}
with  $\omega  \in \mathbb R^4$ and 
\begin{equation}
    S =\begin{bmatrix} S_1 & 0_{2 \times 2}\\0_{2 \times 2} & S_2 \end{bmatrix}, \  S_1 = \begin{bmatrix} 0 & -1\\1 & 0 \end{bmatrix}, \ S_2 = \Omega S_1, \    T_{\nu} = \begin{bmatrix}   0_{3 \times 2}
    \begin{matrix} -\frac{k(1-\eta + \Omega^2)}{2\Omega} & 0 \\ 0 & k \\ k\cos{\phi} & k\sin{\phi} 
    \end{matrix}
    \end{bmatrix}, \ T_{\xi} = 
    \begin{bmatrix}       \begin{matrix} 2 & 0 
    \\
    0 & 0 \\ 0 & -2  \\ -\mu(1-\mu) & 0\end{matrix} & 0_{4 \times 2} \end{bmatrix}e.
\end{equation}
The so-defined exosystem dynamically generates, when setting the initial condition 
$   \omega_0 = \begin{pmatrix}
    \cos{\phi} & \sin{\phi} & \cos{\phi} & \sin{\phi}
    \end{pmatrix}^\top
$, the external signals feeding (\ref{ssmodel}) and, in particular, the quasi Halo orbits \eqref{ref}, the reference trajectories to track, and the periodic eccentricity perturbations \eqref{eq:xi} to asymptotically reject.

As the intuition suggests, the exosystem satisfies the internal model principle \cite{internal_model} so that the problem reduces to stabilizing the internally admissible trajectory (\ref{eq:nu}) while negating the effect of the eccentricity perturbations (\ref{eq:xi}); namely, one seeks for a control law ensuring the regulation error falls to zero asymptotically, i.e.,
\begin{align*}
        e(t) =  Cq(t) - T_\nu \omega(t)\to 0 \text{ as } t \to \infty .
\end{align*}
This argument is at the basis of the controller proposed in \cite{Paolo} deduced by assuming $\xi(t) \approx 0$ and approximating (\ref{ssmodel}) as
\begin{equation}\label{flcontrolmodel}
\begin{aligned}
    \dot q & =  f_0(q) + P(q)\xi + Bu
\end{aligned}
\end{equation}
with
\begin{equation*}
\begin{aligned}
   f_0(q)&= f(q, 0), \quad    P(q) = \frac{ \partial f}{\partial \xi}\Big|_{\xi = 0}.
\end{aligned}
\end{equation*}
In this way, one gets the feedback law 
\begin{equation}\label{ctregfeedback}
    u =  c(\omega) +K(q - \pi(\omega))
\end{equation}
with: $\pi: \mathbb{R}^4 \to \mathbb{R}^6$ and $c: \mathbb{R}^4\to \mathbb{R}^3$, solutions to the Francis-Byrnes-Isidori equations \cite{Francis, Byrnes-Isidori},
\begin{equation*}
\begin{aligned}
    \frac{\partial \pi(\omega)}{\partial \omega}S\omega &= f_0(\pi(\omega)) + P(\pi(\omega))\omega + Bc(\omega)\\
    0 &= C\pi(\omega) - T_\nu\omega ,  
    \end{aligned}
\end{equation*}
 given by
\begin{equation}\label{Pi}
\begin{aligned}
    \pi(\omega) &= \Pi \omega, \quad \Pi = \begin{bmatrix} T_\nu  \quad \\ T_\nu S \end{bmatrix}, \quad 
       c(\omega) = -\begin{bmatrix} 0_{3 \times 3} & I_3\end{bmatrix} \Big( f_0(\pi(\omega)) + P(\pi(\omega))T_\xi\omega \Big) + \Pi S \omega 
\end{aligned}
\end{equation}
with $K$ ensuring $\sigma(A+BK)\subset \mathbb{C}^-$, 
$
    A = \frac{\partial f_0(q)}{\partial q} \Big|_{q = q^\star}
$.
In \cite{Paolo}, this control was shown to provide good results both in terms of quasi-Halo orbit station-keeping error (also when considering higher order terms in \eqref{eq:alpha_beta}) and energy expenditure in the sense of $\| u (t)\|$ in the nominal case of $D(t) = 0$. 
However, such a controller lacks in robusteness with respect to unmodelled uncertainties (e.g., the effect of the SRP) and digital implementation of the control laws as demanded from more recent practical scenarios \cite{NasaReport3}. To make regulation robust with respect to sampling, a solution based on direct digital design has been proposed in \cite{lorenzo} solving the regulation problem on the sampled-data multi-rate equivalent model of (\ref{ssmodel}). Still, in this setting, measures are available only at the sampling periods while the control switches at suitably defined sub-intervals, so making it work in open loop during such sub-intervals. 

\subsection{Multi-rate $L_2$-station under sampling}\label{sampling_background} 
When $D = 0_{3 \times 1}$, the corresponding unperturbed model  (\ref{ssmodel}) possesses a well-defined (vector) relative degree $r = (2 \ 2 \ 2)$; i.e., the following holds:
\begin{itemize}
    \item $ L_{B} h(q) =CB =  0_{3 \times 3} $  while $L_{B}L_f(\cdot, \xi)h(q)  = C \frac{\partial f(q,\xi)}{\partial q} B \not = 0_{3 \times 3}$; 
    \item the decoupling matrix $U(q,\xi) = C \frac{\partial f(q, \xi)}{\partial q} B$
    is non-singular.  
\end{itemize} 
Let us now assume that measurements are available only at $t = k\delta$ (i.e., the sampling instants) with $k\geq 0$ and  the control and the eccentricity perturbation are piecewise constant over the sampling period $\delta \geq 0$; i.e., $\xi(t) = \xi(k)$ and $u(t) = u(k)$ for $t \in [k\delta, (k+1)\delta ($. Then, the dynamics are described at all sampling instants by the so-called single-rate (SR) equivalent sampled-data model which takes the form of a map
\begin{equation}\label{sr_sd_model}
\begin{aligned}
q(k+1) &= F^\delta(q(k),\xi(k), u(k)) \\ 
p(k) &= h(q(k)) = C q(k)
\end{aligned}
\end{equation}
with $q(k) = q(k\delta)$ and
\begin{align}\label{Fd}
F^\delta(q, \xi, u) = e^{\delta \mathrm{L}_{f(\cdot, \xi) + Bu}}q = q + \sum_{j > 0}\frac{\delta^j}{j!}(L_{f(\cdot,\xi) + Bu})^jq.
 \end{align}
 It is a matter of computation to verify that, independently on the continuous-time vector relative degree $r$, the relative degree of each output channel of the sampled-data equivalent model (\ref{sr_sd_model}) always falls to $r_{d} = (1\ 1 \ 1)$; namely, one gets 
 \begin{align*}
     \frac{\partial p(k+1)}{\partial u(k)} =& \frac{\delta^{2}}{2!}  \mathrm{L}_B \mathrm{L}_{f(\cdot, \xi)} h(q) + \mathcal{O}(\delta^{3}) =
     \frac{\delta^{2}}{2!}  C \frac{\partial f(q, \xi)}{\partial q} B + \mathcal{O}(\delta^{3}) 
     \neq 0.
 \end{align*}
As a consequence, the sampled dynamics \eqref{sr_sd_model} possesses a further unstable zero dynamics (the so-called sampling zero-dynamics) of dimension $3$. For this reason, inversion-like techniques via single-rate sampling cannot be achieved while guaranteeing stability of the internal dynamics.
This fact induces some obstructions due to the appearance of the sampling zero-dynamics. To overcome this pathology, 
\textit{multirate sampling} has been introduced and developed in a nonlinear context \cite{SMDNC92}.  

In detail, fixing $\bar{\delta} = \frac{\delta}{2}$ and $u(t) = u^i(k), \  t \in [k\delta, k\delta + i\bar{\delta}(, \ i = 1,2$ one gets the multirate (MR) equivalent model of order $2$ of the form
\begin{equation}\label{mr_sd_model}
	\begin{aligned}
	q(k+1) =& F_{2}^{\delta}(q(k),  \xi(k), \underline{u}(k)) 
	\end{aligned}
	\end{equation}
	with
	\begin{equation*}
	\begin{aligned}
	 F_{2}^{ \delta}(q,  \xi, \underline{u}) =  F^{\bar \delta}(\cdot , \xi, u^2) \circ F^{\bar \delta}(q, \xi, u^1) 
	=      \sum_{j_1, j_2 \geq 0}\frac{\bar \delta^{j_1 +j_2}}{j_1!j_2!} (L_{f(\cdot,\xi) + Bu^1})^{j_1} \circ (L_{f(\cdot,\xi) + Bu^2})^{j_2} q.
	\end{aligned}
	\end{equation*}
At this point, the multi-rate design model used for regulation is deduced by approximating (\ref{mr_sd_model}) as
\begin{align}
    q(k+1) =& q(k) + \frac{\delta}{2}\big(  f^{\delta}(q(k), \omega(k)) + g^{\delta}( \omega(k) )\underline{u}(k)\big)
\end{align}
with
\begin{align*}
    f^{\delta}(q, \omega) =& f(q,  T_\xi \omega) + f(q, T_\xi e^{\frac{\delta}{2} S} \omega ) + \frac{\delta}{2}\frac{\partial f}{\partial q}(q,T_\xi e^{\frac{\delta}{2}S} \omega )f(q, T_\xi \omega)\\
    g^{\delta}( \omega) =& \begin{bmatrix}
    \big( I  + \frac{\delta}{2} \frac{\partial f}{\partial q}(q,T_\xi e^{\frac{\delta}{2} S} \omega ) \big) B &  B
    \end{bmatrix}
    \end{align*}
    where we have used the exosystem (\ref{exosystem}) for predicting the disturbance at the small sampling instants $t = k\delta + \bar \delta$; namely, $\xi(k\delta + \bar \delta) = T_\xi e^{\frac{\delta}{2} S} \omega(k)$ and $\omega(k) = \omega(k\delta)$.

Because $  g^{\delta}( \omega)$ is non-singular the regulation problem is solved by the feedback
 \begin{equation*}\label{mrreg_feedback}
      \underline{u} = \big( g^{\delta}( \omega)\big)^{-1} \Big( A_{d}(q-\pi(\omega)) - f^{\delta}(q,\omega) + \pi(e^{ \delta  S} \omega)\Big)
 \end{equation*}
 with $\pi(\omega)$ as in (\ref{Pi}) and $A_d$ any Schur stable matrix (i.e., with $ \sigma(I + \bar \delta A_d) \in \mathbb{S}^1$). In this sense a MR feedback was used to solve Problem \ref{problem} in an approximate sense. This feedback was shown to give better results  when applied to the simulation model (\ref{ssmodel}) without SRP compared to the emulation-based implementation of the continuous-time regulation feedback \eqref{ctregfeedback} (i.e., when directly implemented via zero-order-holding (ZOH) and no redesign). 
 On the other hand, a known limitation of MR sampled-data control is the lack of robustness. 

\subsection{$L_2$-station keeping under nonlinear MPC}\label{sec:NMPC_recalls}


With reference to the SR sampled-data model (\ref{sr_sd_model}) and assuming $\xi = 0_{4 \times 1}$, a station-keeping nonlinear MPC has been proposed in \cite{polynMPC} based on the online solution, at all sampling instants, of a finite horizon optimal control problem of the form;
\begin{subequations}\label{generalMPC}
    \begin{align}
    & \min \big{\{ }   V_{n_p}(q(k+n_p)) +  \sum_{i=1}^{n_p-1}  \|p(k+i)- \nu(k+i)\|_Q^2 + \|u(k+i-1)\|^2_{R} \big{ \} } \label{stage_cost} \\ 
     \text{s. t.}\quad 
     &q(k + 1) = F^{\delta}(q(k),0, u(k)) \label{prediction}\\
     &q(k + i)\in \mathcal{X} , \quad  u(k+j)  \in \mathcal{U}, \quad i = 1 \hdots, n_p-1, j = 0, \hdots, n_c-1 \label{constraints}\\
     &q(k+n_p) \in \mathcal{X}_{n_p} \label{terminalSet}
    \end{align}
\end{subequations}
 where: the sampled-data equivalent model is used as prediction model \cite{grune}; $\nu(k) = \nu(k\delta)$ is the sample of the reference trajectory in (\ref{ref});  $Q\succeq 0$ and $R\succ 0$ are the penalizing weights; $n_p, n_c$ are the prediction and control horizons;  $\mathcal{X}, \mathcal{U}$ are the states and control constraints sets assumed, as usual, compact, convex and containing the origin;  $V_{n_p}$ and $\mathcal{X}_{n_p}$ represent the terminal cost and terminal constraint set respectively. Those terminal ingredients are incorporated into the optimization problem (\ref{generalMPC}) for providing  closed-loop stability and convergence \cite{Rawlings}. Instability in closed loop, when no further constraint is included, arises due to the generally unstable zero-dynamics induced by single-rate sampling \cite{SMDNC86}.
 
\begin{remark}
In Halo-orbit station-keeping applications usually one has box constraints on the controls \cite{survey}.
\end{remark}

The nonconvex \emph{nonlinear programming} (NLP) problem (\ref{generalMPC}) can be recast into a polynomial optimization problem  \cite{SOS,polynMPC} of the form
\begin{subequations}\label{polynmpc}
\begin{align}
    V^\star &= \max_{U \in \mathcal{U}, \lambda \in \mathbb{R}} \lambda\\
 &s.t \quad p_0(U) - \lambda \geq 0\label{polynmpc_constr}
\end{align}
\end{subequations}
with $\lambda \in \mathbb R$, $U = (u(k), \ \hdots \, \ u(k+n_p-1))^\top$ being the decision variable, and denoting the single-shooting prediction of the output for $i$ steps by $H_i(U)$;
\begin{align*}
    p_0(U) &= H_1(U) + H_2(U) + \hdots + H_{n_p}(U)\\
    H_1(U) &=  \|\nu(k+1) - CF^\delta(q(k), 0, u(k))\|_Q\\
    H_2(U) &=  \|\nu(k+2) - C F^\delta(\cdot,0 , u(k+1))\circ F^\delta(q(k),0, u(k))\|_Q\\
    &\vdots\\
    H_{n_p}(U) &= \|\nu(k+n_p) - C F^\delta(\cdot,0 , u(k+n_p-1))\circ \dots \circ F^\delta(q(k),0, u(k))\|_Q
\end{align*}
so to guarantee, if any, convergence to a solution in finite time and the corresponding global optimality.
The above problem is a Sum-Of-Squares (SOS) polynomial minimization problem \cite{Putinar} subject to convex polynomial constraints, for which one can utilize a Semi-Definite Programming (SDP) solver to reach global optimal solutions in polynomial time. However, as noted in \cite{polynMPC}, when solving (\ref{polynmpc}) via SDP solvers two issues arise: semi-definite programming solutions to (\ref{polynmpc}) are computationally demanding (depending on the degree of the sum of squares polynomial relaxations for $p_0(U)-\lambda$), and the closed loop system may lack in robustness since the employed prediction model neglects both eccentricity and SRP and for which no reference signal pre-processing is done.

In the context of tracking  known limitations with nonlinear MPC are related to two main aspects: recursive feasibility of the optimization problem that is linked to admissibility of the reference trajectory for the dynamics, and  boundedness of the closed-loop trajectories \cite{AdmissibleSet}. %

\section{Station-keeping under sampled-data model predictive control}\label{SEC:IV}

To solve issues arising in both the recalled sampled-data approaches (i.e., nonlinear multi-rate control and standard MPC), we propose to integrate a Multi-Rate (MR) sampled-data planner in MPC while reducing the complexity of the optimization problem and removing terminal costs and further fictitious constraints. 
 The new control scheme combines nonlinear regulation under multi-rate sampling and single-rate MPC so to simplify the optimization problem associated to the MPC and, at the same time, improving robustness of the closed loop.

With reference to Figure \ref{fig:scheme}  below, we detail the proposed single-rate sampled-data quasi Halo orbit station-keeping control scheme. This proposed scheme solves Problem \ref{problem} combining an outer trajectory planner together with an inner nonlinear MPC controller working at different sampling rates. More In detail, the trajectory is planned at all $t = 2 k\bar \delta$ (the so-called planning instants), whereas the control is computed by the MPC inner controller at all $t = k\bar \delta$ (the sampling instants). The control is composed by the \textbf{multi-rate reference governor} (or planner) and the \textbf{inner-loop MPC controller}. More precisely, given samples of the quasi Halo orbit reference generated by the exosystem (\ref{eq:nu}), the MR reference governor produces a sequence of admissible position trajectories, denoted by $\hat p^i(k) = \hat p(k+i)$, based on a simplified model of the spacecraft under the regulation feedback (\ref{ctregfeedback}) when neglecting the SRP;  such a reference is then fed to a simplified nonlinear MPC control problem, thus ensuring tracking, while mitigating the effect of the unmodelled additive SRP disturbance.

\begin{figure}[!h]
    \centering
    \includegraphics[width=0.8\textwidth]{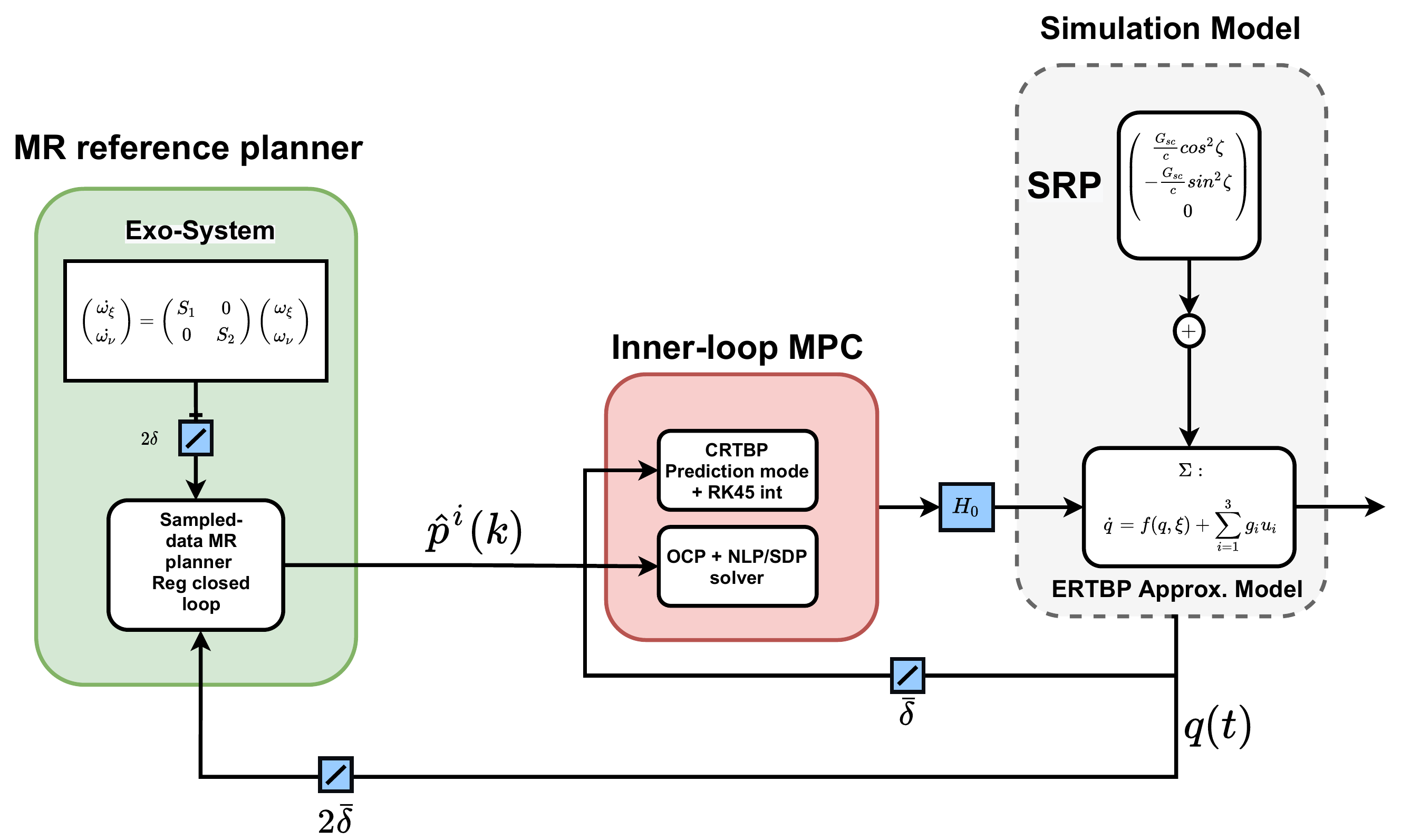}
    \caption{The proposed control scheme MR MPC}
    \label{fig:scheme}
\end{figure}

\subsection{MR as a reference governor} 

Starting from the design in Section \ref{mrreg_feedback}, we consider the feedback system (\ref{ssmodel}) under the control  (\ref{ctregfeedback}) computed for $K = 0_{3 \times 6}$ and $D = 0_{3 \times 1}$; namely, we get 
\begin{align}\label{closed_loop_simp}
    \dot q(t) = \tilde f(q(t), \omega(t)) + B \tilde u_i(k)
\end{align}
for $t \in [(k+\frac{i-1}{2} \delta, (k+\frac{i}{2})\delta ( $ and $i = 1, 2$ and $\tilde f(q, \omega) = f(q, T_\xi \omega) + Bc(\omega)$. Accordingly, the corresponding MR sampled-data equivalent model is approximately provided by
\begin{equation}\label{MRplanner}
    \begin{aligned}
        q(k+1) &= \tilde F^{\delta}_2(q(k), \omega(k), \underline{\tilde u}(k)) 
    \end{aligned}
\end{equation}
with 
$\underline{\tilde u} = (u_1^\top \ u_2^\top )^\top \in \mathbb{R}^6 $, the map
\begin{align*}
     \tilde F^{\delta}_2(q, \omega, \underline{\tilde u}) = q +  \frac{\delta}{2} \big(\tilde f^{\delta}(q,\omega) + \tilde g^{ \delta}(\omega) \underline{\tilde u} \Big)
\end{align*}
\begin{align*}
    \tilde f^{\delta}(q,\omega) &= \Big( I + \frac{\delta}{4} (A + \frac{\partial \Gamma}{q}(q,T_\xi e^{ \frac{\delta}{2} S}\omega) )\Big) Aq, \quad 
       \tilde g^{ \delta}(\omega) = \begin{bmatrix} B + \frac{\delta}{4}AB & \frac{ \delta}{4}\frac{\partial \Gamma}{q}(q,T_\xi e^{ \frac{\delta}{2} S}\omega)B \end{bmatrix}\\
     \Gamma(q, \omega) &= - M (p+T_\nu \omega) - 2N(\dot p + T_\nu S \omega) -(2Mp+2N\dot p)(e_1T_{\xi}\omega) - Mp(e_2T_{\xi}\omega) - N\dot p (e_3T_{\xi}\omega) \\&-  (1-\mu)[\frac{p - \tilde d_1}{\|p - \tilde d_1\|^3} + \frac{T_\nu \omega - \tilde d_1}{\|T_\nu\omega - \tilde d_1\|^3}] - \mu [\frac{p - \tilde d_2}{\|p - \tilde d_2\|^3} + \frac{T_\nu \omega - \tilde d_2}{\|T_\nu \omega - \tilde d_2\|^3}] + P(\pi(\omega))T_\xi\omega - T_\nu \omega
\end{align*}
\begin{align*}
    \tilde d_1 &= \begin{bmatrix} -\mu - \frac{e_4T_\xi\omega_\xi}{1-\mu} & 0 & 0 \end{bmatrix}^\top , \quad \tilde d_2 = \begin{bmatrix} 1-\mu + \frac{e_4T_\xi\omega_\xi}{\mu} & 0 & 0 \end{bmatrix}^\top, \quad A = \begin{bmatrix} 0_{3 \times 3}  & I_3 \\ 0_{3 \times 3} & 0_{3 \times 3} \end{bmatrix}
\end{align*}
Accordingly, station keeping is guaranteed for the discrete-time model (\ref{MRplanner}) by the planned feedback law  
\begin{align}\label{mrplannedControls}
    \underline{\tilde u}(q, \omega) &=  \big(\tilde g^{\delta}( \omega)\big)^{-1}\Big( A_d (q-\pi(\omega)) + \pi(e^{\delta S}\omega) - \tilde f^{\delta}(q, \omega))\Big) 
\end{align}
and $A_d$ being such that  $
    \sigma(I+\delta A_d) \subset \mathbb{S}^1
$ for all $\delta >0$.

\begin{remark}
It can be verified that  $\Gamma (q, \omega)$  vanishes as $p \to T_\nu \omega$ that is, as the satellite reaches the Halo orbit, apart from the external perturbation. More in detail, one computes at the Halo orbit (i.e., $p = T_\nu \omega$) $$\Gamma(q, \omega) = -(e_1T_{\xi}\omega_\xi)(2Mp+2N\dot p) - (e_2T_{\xi}\omega_\xi)Mp - (e_3T_{\xi}\omega_\xi)N\dot p  + P(\pi(\omega))T_\xi\omega_\xi = \bar{\Gamma}(q)\omega_\xi.$$  Moreover, denoting $\omega_\xi = \begin{bmatrix}
I_2 & 0_{2 \times 2}
\end{bmatrix} \omega$, one gets $\|\bar\Gamma(q) \omega_\xi\|$ is bounded at the quasi Halo orbit since $\| \omega_\xi\| \leq c, c \in (0 , 1]$.
\end{remark}

At all planning instants $t = k\delta$, the simplified MR model (\ref{MRplanner}) under the feedback (\ref{mrplannedControls}) provides an admissible sequence of bounded references $\hat p^{i}(k) = \hat p(k\delta+ \frac{i \delta}{2})$ with $k = 1, \dots, \hat n_p$, $ i = 1, 2$ for the original dynamics (\ref{ssmodel}) over the prediction horizon of length $n_p = 2\hat n_p $ and $\hat n_p >1$.
More rigorously, for $i = 1,2$, $k = 1, \dots, \hat n_p$ and $q^0(k) = q(k\delta)$ the multi-rate planner gets the form
\begin{subequations}\label{sr_planned_traj}
\begin{align}
     q^{i}(k) =& \tilde F^{\bar \delta}(q^{i-1}(k), \tilde u_i(k))\\
    \hat p^{i}(k) =& C q^{i}(k)
\end{align}
\end{subequations}
with the single-rate map associated to (\ref{closed_loop_simp}) given by
\begin{align*}
    \tilde F^{\bar \delta}(q, \tilde u) = q + \bar \delta\Big( \tilde f(q, \omega) + B \tilde u \Big) + O(\bar \delta^2)
\end{align*}
Accordingly, such references can be directly fed to MPC controller for which, then, the optimization problem can be notably simplified with respect to the one recalled in Section \ref{sec:NMPC_recalls}, typically used in the literature, as the following section details. 

\subsection{The inner-loop MPC controller}

Assuming now the references computed by the planner (\ref{sr_planned_traj}), the actual single-rate control is computed via MPC solving at all sampling instants $t = k\bar \delta$ the modified optimization problem (\ref{generalMPC}) of the form
\begin{subequations}\label{MPC_pb}
\begin{align}\label{cost}
V^\star &= \min_{U} \big\{ \sum_{\ell = 0}^{ \hat n_p-1} \sum_{i = 1}^{2 }\| p((2k+\ell + i) \bar \delta ) - \hat p^i((2k+\ell)\bar \delta) \|_{Q} + \| u(2k+\ell + i-1))\|_R  \big\} \\
\label{dynamicsConstraints} 
  \text{ s. t. } \quad  & q(k+1) = F^{\delta}(q(k),0,u(k)) \\
 & lb \leq u(k+i-1) \leq ub ,\quad i = 1, \hdots, n_p\label{controlConst} \text{ and } n_p = 2\hat n_p.
\end{align}
\end{subequations}
Some comments are in order.
\begin{itemize}
 \item  The MPC problem we propose is notably simplified with respect to the one typically adopted in the literature. As a matter of fact, the boundary conditions (\ref{terminalSet}) and the terminal penalty $V_{n_p}(\cdot)$ are no longer needed. This is due to the fact that the references $\hat p^i(\cdot)$ provide bounded and admissible trajectories for the dynamics by construction of the multi-rate planner (\ref{MRplanner}). 
  This is specially true for $R$ small enough, when neglecting (\ref{controlConst}) \cite{ElobaidIFAC}. In this sense, the admissible references $\hat p^i(k)$ are specifically designed to ensure better performances while simplifying the optimization problem.  The generated references depend on the quasi Halo references obtained from (\ref{eq:nu}). 

  \item  Expression (\ref{dynamicsConstraints}) represents a simplified prediction model for MPC being an approximate sampled-data equivalent CRTBP model.  

 \item  The constraint (\ref{controlConst}) takes into account possible saturations and physical limitations of the actuators and safety margins on thrusts. Note that (\ref{controlConst}) are simple algebraic convex box constraints. Consequently, polynomial optimization reformulation of the nonlinear MPC problem above is possible replacing in (\ref{polynmpc}) the constraint $U \in \mathcal{U}$, with $U \in \mathcal{L} = \{ U \in \mathbb{R}^{3n_p} : L [U \ I]^\top \leq 0 \}$ where
\begin{align}\label{polyBox_constraints}
 L &=  \begin{bmatrix*} I & 0 & \hdots & 0 &-ub\\ I & 0 & \hdots & 0 &lb \\ \vdots & \vdots & \ddots & \vdots & \vdots\\I & 0 & \hdots & 0 &-ub\\ I & 0 & \hdots & 0 &lb \end{bmatrix*} \in \mathbb{R}^{3n_p \times 3n_p + 3 }
\end{align}

    \end{itemize}

To sum up, the control scheme we propose is summarized in the algorithm below.

\begin{algorithm}[H]
\SetAlgoLined
\KwResult{Solving Problem \ref{problem}: $p(k) \to \nu(k)$}
 \small{initialization \;
$q(k) \leftarrow q(t = k\delta), Q \leftarrow Q, R \leftarrow R, ub \leftarrow u_{MAX}, lb \leftarrow u_{MIN} $\;}
 \While{$t \geq 0$}{
  \If {$t = (k+j)\delta, j \in \mathbb{Z}_{\geq 0}$}{
  $k \leftarrow k+j$\; 
  $(\hat{p}^i)=$ Planning($q(k), T_\nu \omega(k)$)\;
  $u(k)=$ SolveNMPC() \;
  }
  \For{$t = k\delta+i\bar\delta, i = 1, 2$}{
   $u(k)=$ SolveNMPC()\;
   }
 }
 \caption{planning and control}
\end{algorithm}

\begin{remark}
In the algorithm above, the \textit{SolveNMPC()} routine is a NLP (\textcolor{black}{ equivalently SDP}) routine that solves the optimization problem \ref{MPC_pb}. This can be any off the shelf routine. The implementation details and a brief discussion on the computational aspects associated with this routine are deferred to the next section. The $\text{if}$ statement in the algorithm specifies that planning and control are carried at different sampling rates as already discussed. This is precisely why we need to propagate the MR planner (\ref{MRplanner}) and compute the admissible reference $\hat p^i(k)$  over double the prediction horizon.
\end{remark}

The proposed control scheme  solves the digital $L_2$ station-keeping problem while addressing the limitations discussed in Section \ref{SEC:III}. It makes use of a simplified MR model to provide admissible references for a nonlinear MPC with an improved prediction model.
The corresponding advantages will be shown in the next section through simulations depicting various realistic scenarios. 


\section{Simulations and comparative discussions}\label{SEC:V}


A comparative study is carried out making use of Matlab/Simulink$\copyright$.The proposed control scheme will be denoted by MR MPC and compared to nonlinear MPC via polynomial optimization in \cite{polynMPC, survey} (referred to as PolyNMPC), nonlinear regulation and feedback linearization (FL) as proposed in \cite{Paolo} . 
For PolyNMPC, a polynomial approximation of the CRTBP model is used for prediction deduced from Runge–Kutta (RK) numerical integrator of (\ref{ssmodel}) assuming $D(t) = 0_{3 \times 1}$ and $\xi = 0_{4 \times 1}$. Consequently the nonlinear MPC problem (\ref{MPC_pb}) is transformed into a polynomial optimization problem using sum of squares (SOS), as detailed in (\ref{polynmpc}). This equivalent optimization problem is then solved in a receding horizon approach utilizing SOSTools$\copyright$ \cite{sostools}.

We assume as simulation time $T \approx 2.7 days = m \bar \delta$ coinciding the time needed for at least two revolutions around the orbit and $\bar \delta >0$. The comparison is aimed at  showing how the various controllers succeed in keeping the spacecraft on the Halo orbit based, also, on Key Performance Indicators (KPIs) below.

\textbf{Tracking Error} in terms of the root-mean-square value of the spatial tracking error; namely,  $e_{RMS}(t) = \sqrt{\frac{1}{3}(\| \nu(t) - p(t)\|)}$ and $e_{RMS}(T) = \sqrt{\frac{1}{3}(\| \nu(T) - p(T)\|)}$ in non-dimensional units.  %

\textbf{Energy Expenditure} based on the $\mathcal{L}_2$-norm of the controls for the duration of the flight (i.e., $T$); i.e.,  $ EE_T = \sqrt{\int_0^T \|u(t)\|^2 \mathrm{d}t } = \bar \delta \sqrt{  \sum_{k = 0}^{m} \| u(k)\|^2}$ with $u(k) = u(k\bar \delta)$ and $k\geq 0$ in $m/s^2$. Additionally, the instantaneous control norm (i.e., $\|u(t) \| $ )  is  reported as well as an indicator of the so-called \textit{velocity budget}.


More in details, starting from several initial conditions (e.g. near $L_2$, close to the quasi-Halo orbit), the proposed MR MPC scheme is compared with different control strategies available in the literature in terms of robustness with respect to the effect of:

\begin{itemize}
\item \textbf{unmodelled perturbations and actuator saturations} (Sections \ref{scenario1}-\ref{scenario3}) when the spacecraft starts from various initial conditions (e.g. near $L_2$, close to the quasi-Halo orbit) in nominal conditions with no SRP terms and saturation on thrusts, and when assuming SRP terms and saturation of the actuators which were neglected during the design.

\item \textbf{sampling} (Section \ref{scenario4}) in presence of perturbations but no limits on the thrusts. 
\end{itemize}
The computational aspects and the application of a Real Time Iteration MPC (RTI MPC) solver to the proposed control scheme conclude the comparative study (Section \ref{scenario5}).

\begin{remark}
The first two simulation scenarios (which are not realistic in general) are carried out for allowing the comparison of the proposed controller with previous works in the literature reporting the same scenario.
As a matter of fact, the point $L_2$ exists only in lower fidelity  models such as the restricted three body problem. Once external forces are included and the fidelity of the model improved, such a point may not even exist.  
\end{remark}

\begin{remark}
The penalizing weighing matrices are fixed throughout the simulation scenarios as $Q = diag\{10,10,10,1,1,1\}$, $ \ R = 0.1 I_3$. The small penalty on the control is chosen so that mimicking the cheap control setting one obtains, in nominal conditions, an almost dynamic inversion-like behaviour \cite{ElobaidIFAC}. These values maybe widely different from, for example, the ones reported in \cite{polynMPC}. This is accounted for by the fact that weighing matrices cannot be used to infer the performances achieved when different constraints, reference signal pre-processing and optimization solvers are used. For instance, the work \cite{polynMPC} dealt with station-keeping around $L_1$ in an unperturbed setting using a slightly different prediction model.
\end{remark}

\subsection{First comparison starting from $L_2$ in nominal conditions}\label{scenario1}
\begin{figure}[!t]
\centering
\includegraphics[width=8.5cm]{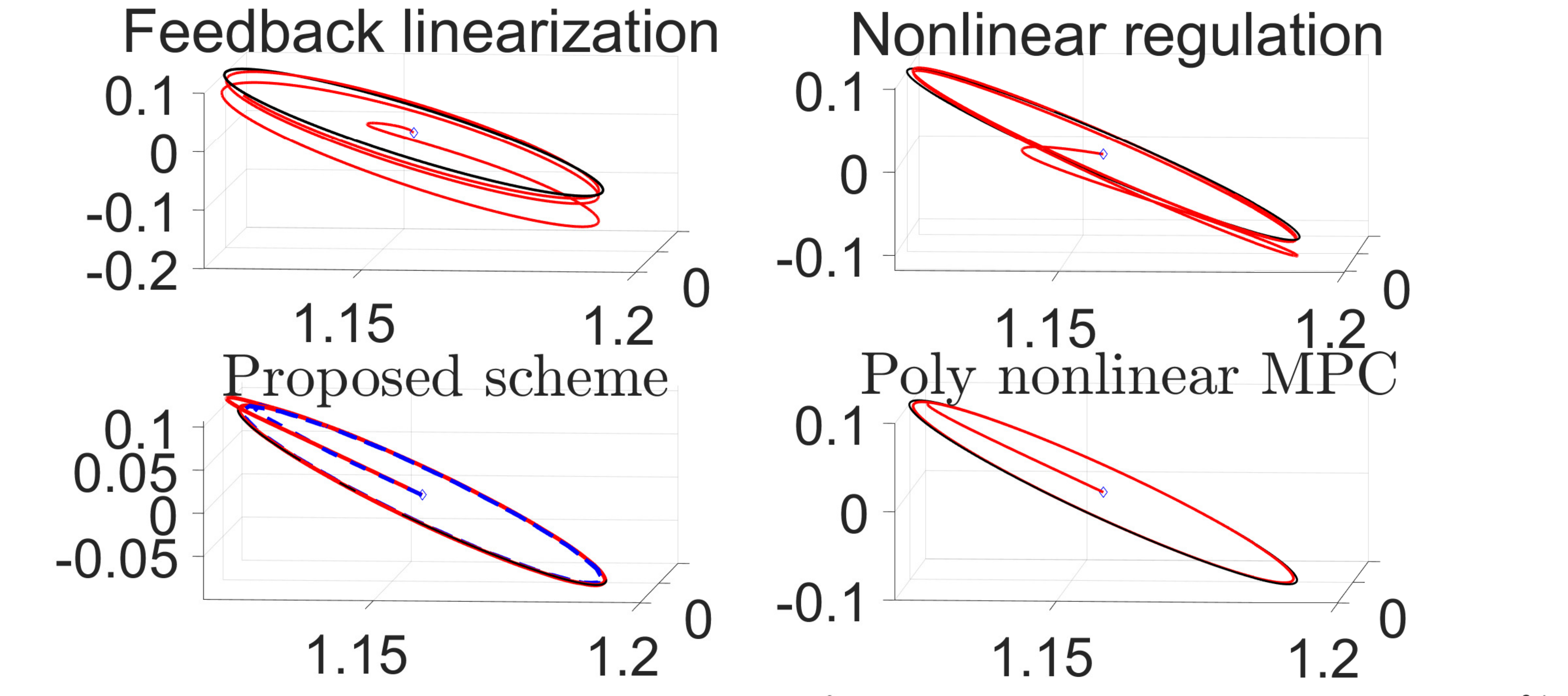}
\caption{Comparison of control strategies. Red specifies the controlled trajectory in $\mathbb R^3$. Black the nominal approximate halo reference, and dashed blue the MR planned reference. Plots are in non-dimensional units.}
\label{sim_1comparisons}
\end{figure}

\begin{figure}[!t]
\centering
\includegraphics[width=8.5cm]{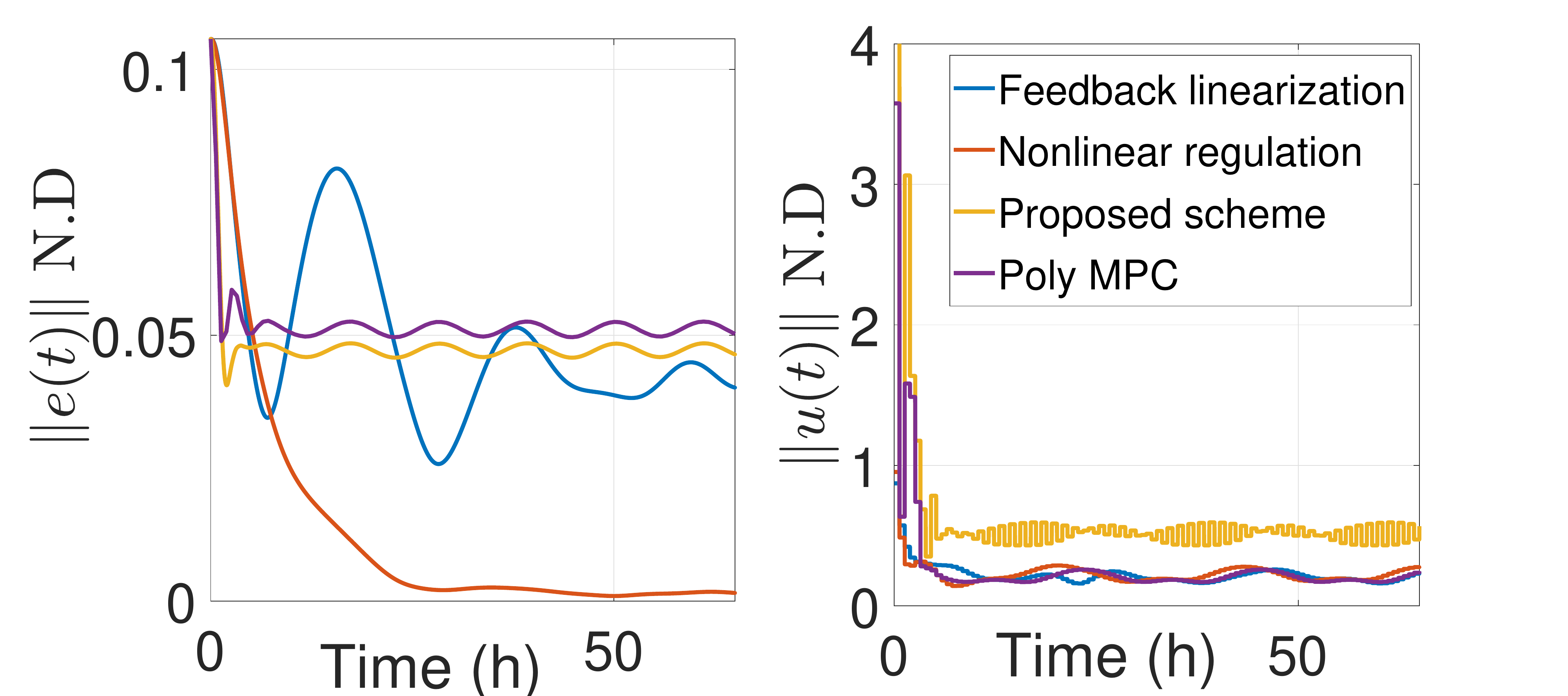}
\caption{Comparison of performances of the various control strategies in the nominal unperturbed case.}
\label{sim_1comparisons_performance}
\end{figure}

\begin{figure}[!t]
\centering
\includegraphics[width=8.8cm]{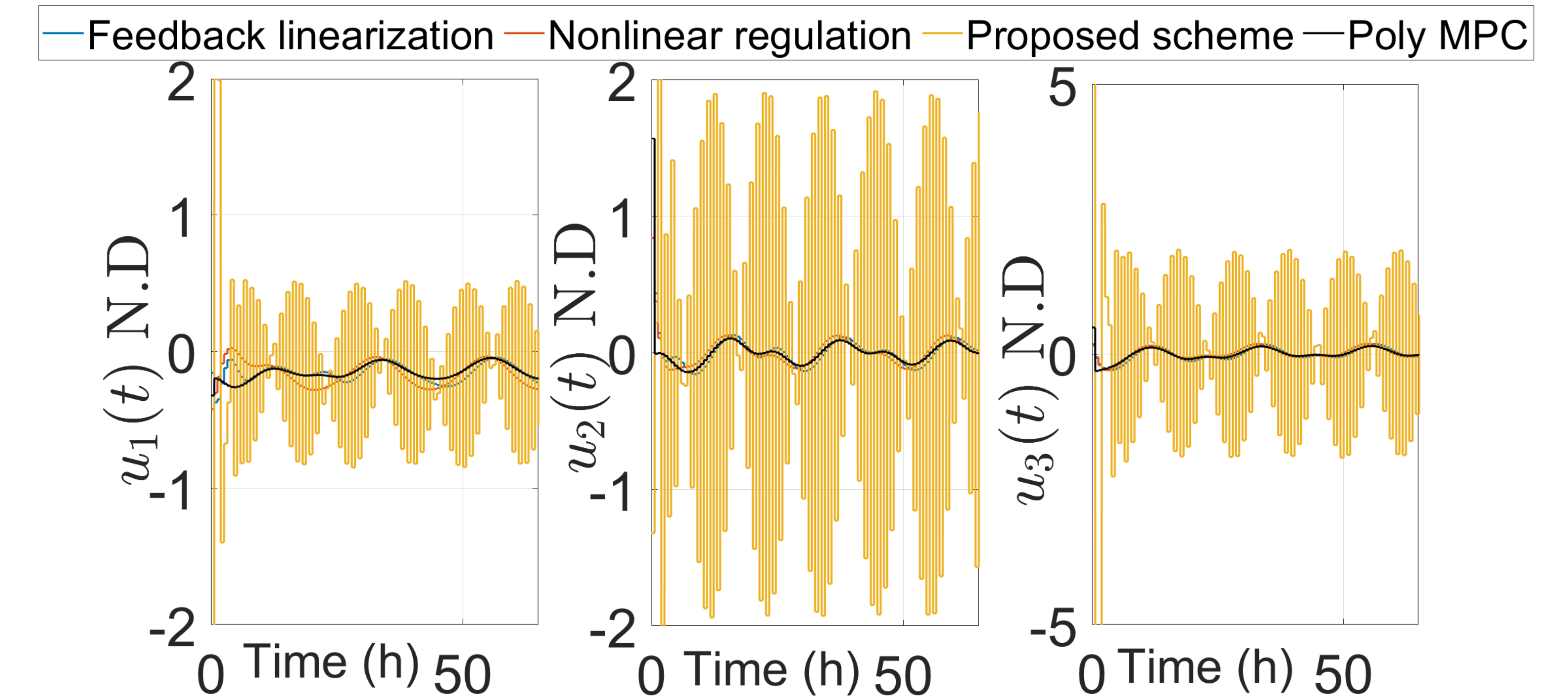}
\caption{Comparison of control inputs of the various control strategies in the nominal unperturbed case.}
\label{sim_1comparisons_controls}
\end{figure}

The spacecraft is assumed to start in the translunar point i.e. $q(0) =  (
L_2 \ 0 \ 0 \ 0 \ 0 \ 0
)^\top$ in the rotating frame with no SRP terms. The proposed MR MPC and the PolyMPC are set with the same penalizing weights to facilitate comparison. The sampling period $\bar \delta =  0.65 \text{hr}$ in dimensional time (this is a reasonable sampling period for this application see Sec. 4.A.7 in \cite{survey}). Figure \ref{sim_1comparisons} shows the controlled trajectories in $\mathbb{R}^3$ in non-dimensional units under the aformentioned controllers. The results show that all the control schemes are comparable. This is further confirmed in Figure \ref{sim_1comparisons_performance}, where clearly nonlinear regulation outperforms the rest both in terms of tracking error and energy expenditure. It is worth mentioning that the proposed scheme outperforms its Poly MPC counterpart in terms of tracking error at the cost of higher control effort (see also Figure \ref{sim_1comparisons_controls}). Table \ref{table_1} summarizes the main KPIs and performances of the various controllers tested in this scenario.

\begin{table}[ht]
\centering
\caption{Summary of main KPI comparisons in scenario 1 at $T = 65 \text{hr}$ }
\begin{tabular}{||c c c||} 
 \hline
 &   $e_{RMS}(T)$  N.D  &  $EE_T$ N.D \\ [0.4ex] 
 \hline\hline
 FL   & 0.04  & 2.238 \\
 Regulation  & 0.002  & 2.29\\
 MR MPC  & 0.047  & 1.927 \\
 PolyMPC  & 0.069  & 0.98\\
 \hline \hline
 \end{tabular}
\label{table_1}
\end{table}

\subsection{Second comparison starting from $L_2$ with SRP and control limits}\label{scenario2}

\begin{figure}[!t]
\centering
\includegraphics[width=8.5cm]{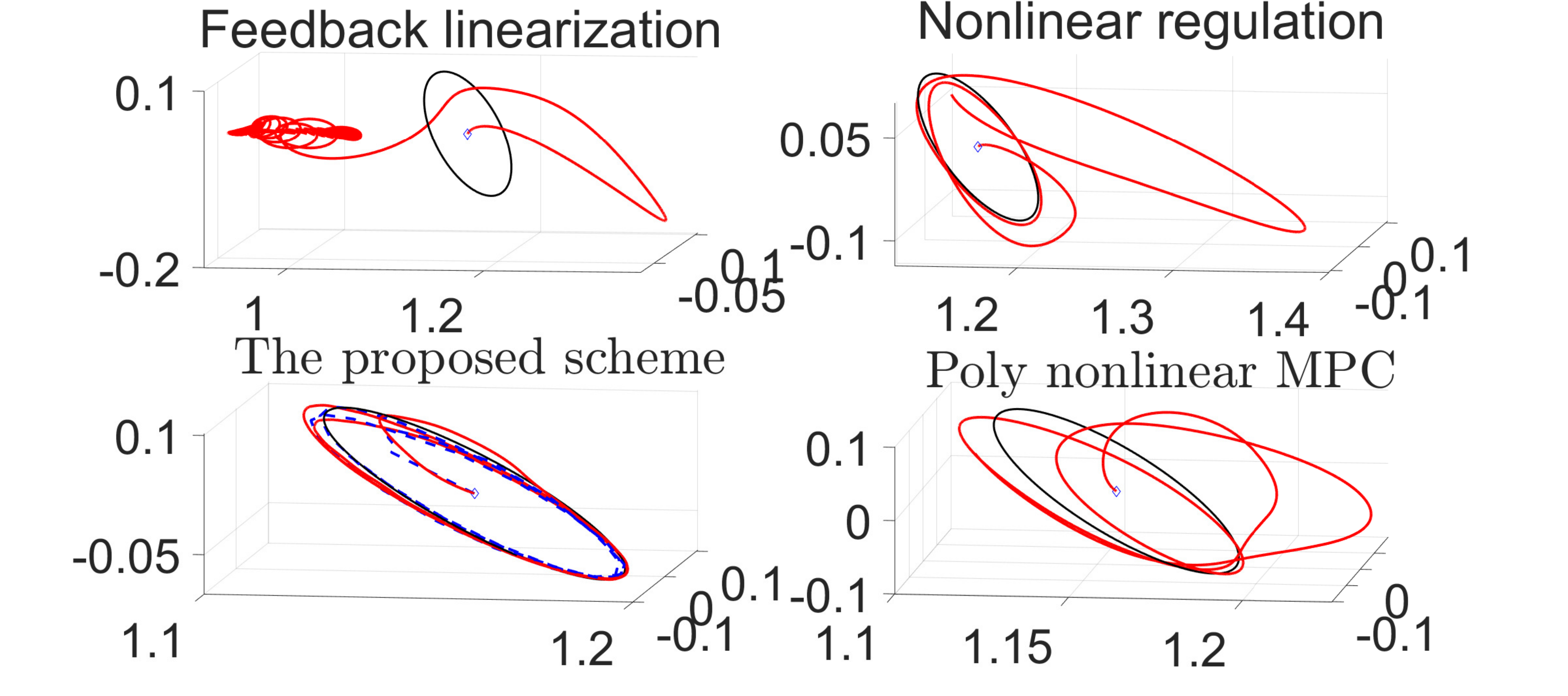}
\caption{Comparison of control strategies. Red specifies the controlled trajectory in $\mathbb R^3$. Black the nominal approximate halo  reference, and dashed blue the MR planned reference. Plots are in non-dimensional units.}
\label{sim_2comparisons}
\end{figure}

\begin{figure}[!t]
\centering
\includegraphics[width=8.8cm]{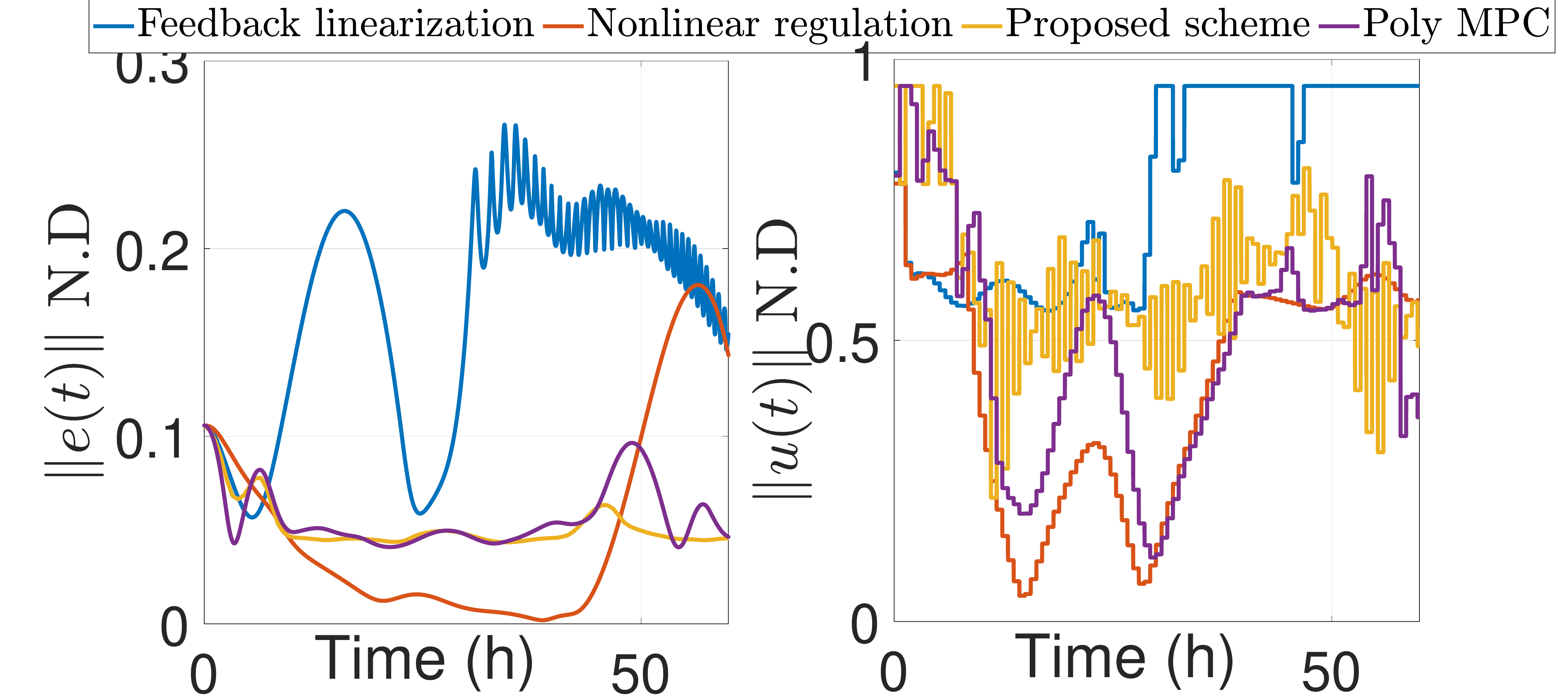}
\caption{Comparison of performances of the various control strategies in the second scenario.}
\label{sim_2comparisons_performance}
\end{figure}

This is a more realistic scenario in which the spacecraft starts at the translunar libration point while being subject to the primaries perturbations, solar radiation pressure as well as assuming each thrust being able to provide a maximum of $0.55\text{[ND]}$. the saturation limit on the thrust in non-dimensional units was chosen mainly to stress-test the compared methodologies. The sampling period is maintained at $\bar \delta = 0.65\text{hr}$. While emulation of the standard feedback linearization fails with this limit on the control (Figure \ref{sim_2comparisons}), emulation of nonlinear regulation feedback yields better results in terms of tracking error and control effort. Intuitively, better performance is obtained when utilizing the proposed MR MPC approach, and PolyMPC (Figure \ref{sim_2comparisons_performance})) being well equipped to handle constraints.
Unlike the previous case, the propose control scheme yields the best tracking performance, even outperforming nonlinear regulation which fails to maintain long-term station-keeping. This is achieved at the expense of a relatively higher control effort over the station-keeping simulation period. Table \ref{table_2} summarizes the main KPIs and performances of the various controllers tested in this scenario.

\begin{figure}[!t]
\centering
\includegraphics[width=8.8cm]{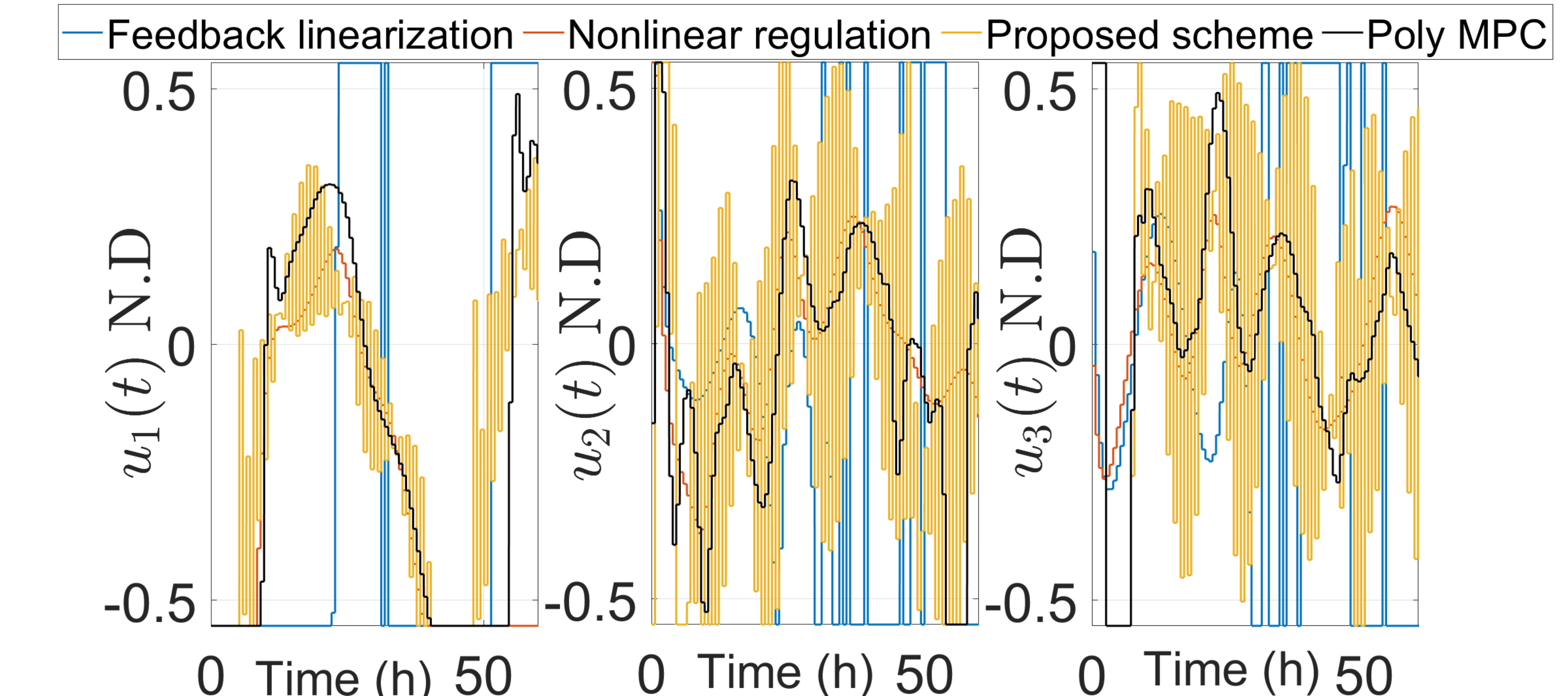}
\caption{Comparison of control inputs of the various control strategies in the second scenario.}
\label{sim_2comparisons_controls}
\end{figure}


\begin{table}[ht]
\centering
\caption{Summary of main KPI comparisons in scenario 1 at $T = 65 \text{hr}$ }
\begin{tabular}{||c c c||} 
 \hline
 &   $e_{RMS}(T)$  N.D  &  $EE_T$ N.D \\ [0.4ex] 
 \hline\hline
 FL   & N/A  & N/A \\
 Regulation  & 0.044  & 4.2001\\
 MR MPC  & 0.0255  &6.0708 \\
 PolyMPC  & 0.0323   & 5.3414\\
 \hline \hline
 \end{tabular}
 \label{table_2}
\end{table}

\subsection{Third comparison starting close to the orbit with SRP and control limits}\label{scenario3}

\begin{figure}[!t]
\centering
\includegraphics[width=8.5cm]{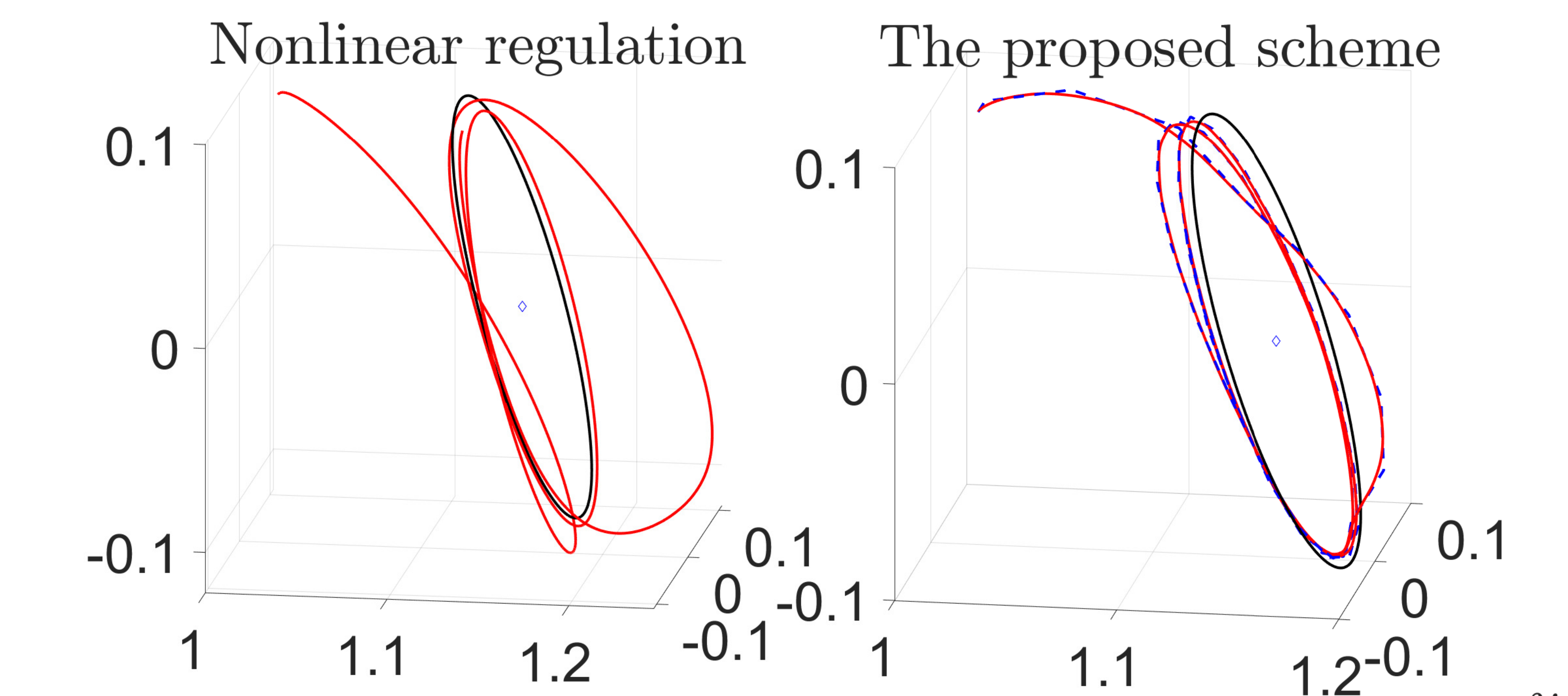}
\caption{Comparison of control strategies. Red specifies the controlled trajectory in $\mathbb R^3$. Black the nominal approximate halo reference, and dashed blue the MR planned reference. Plots are in non-dimensional units.}
\label{sim_3comparisons}
\end{figure}

\begin{figure}[!t]
\centering
\includegraphics[width=8.8cm]{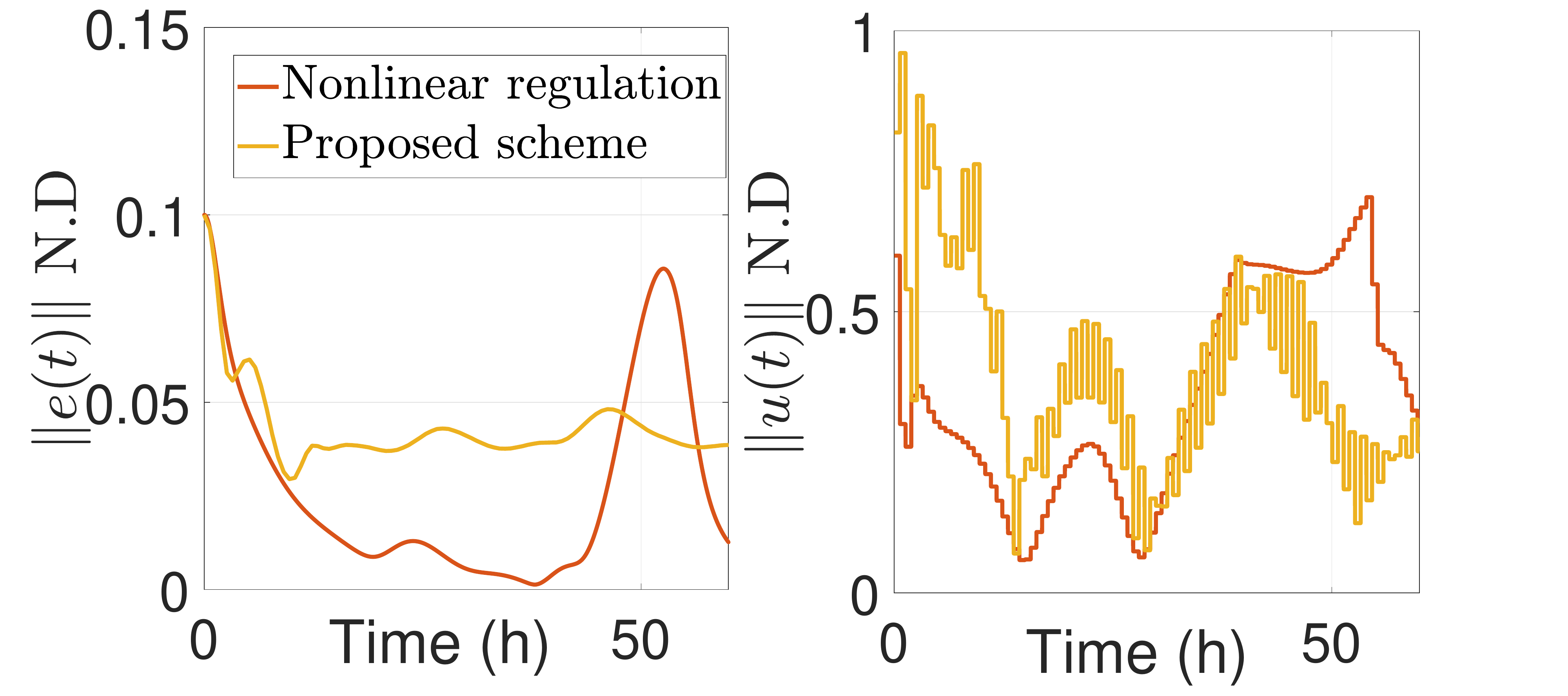}
\caption{Comparison of performances of the various control strategies in the third scenario.}
\label{sim_3comparisons_performance}
\end{figure}

In this typically studied and more realistic scenario consistent with the literature, we start near the orbit with a so-called orbit injection error of $300 \text{km}$ in dimensional units and the same control limit imposed previously. While at the end of simulation period, the nonlinear regulation seems closer to the orbit (Figure \ref{sim_3comparisons_performance}), yet, it is clear from Figure \ref{sim_3comparisons} that the proposed controller maintains the space craft closer to the Halo orbit consistently during the whole simulation period with no jumps in $\|e(t)\|$ value. It is also worth mentioning that in this scenario both the PolyMPC controller and the feedback linearization approach fail to achieve station-keeping (and is the corresponding figure is thus omitted). Table \ref{table_3} reflects the previous discussion.

\begin{table}[ht]
\centering
\caption{Summary of main KPI comparisons in scenario 1 at $T = 65 \text{hr}$ }
\begin{tabular}{||c c c||} 
 \hline
 &   $e_{RMS}(T)$  N.D  &  $EE_T$ N.D \\ [0.4ex] 
 \hline\hline
 FL   & N/A  & N/A \\
 Regulation  & 0.0174  & 5.4\\
 MR MPC  & 0.0253  & 6.5410 \\
 PolyMPC  & N/A   & N/A\\
 \hline \hline
 \end{tabular}
 \label{table_3}
\end{table}



\subsection{Fourth comparison on the effect of $\delta$}\label{scenario4}
\begin{figure}[!t]
\centering
\includegraphics[width=8.5cm]{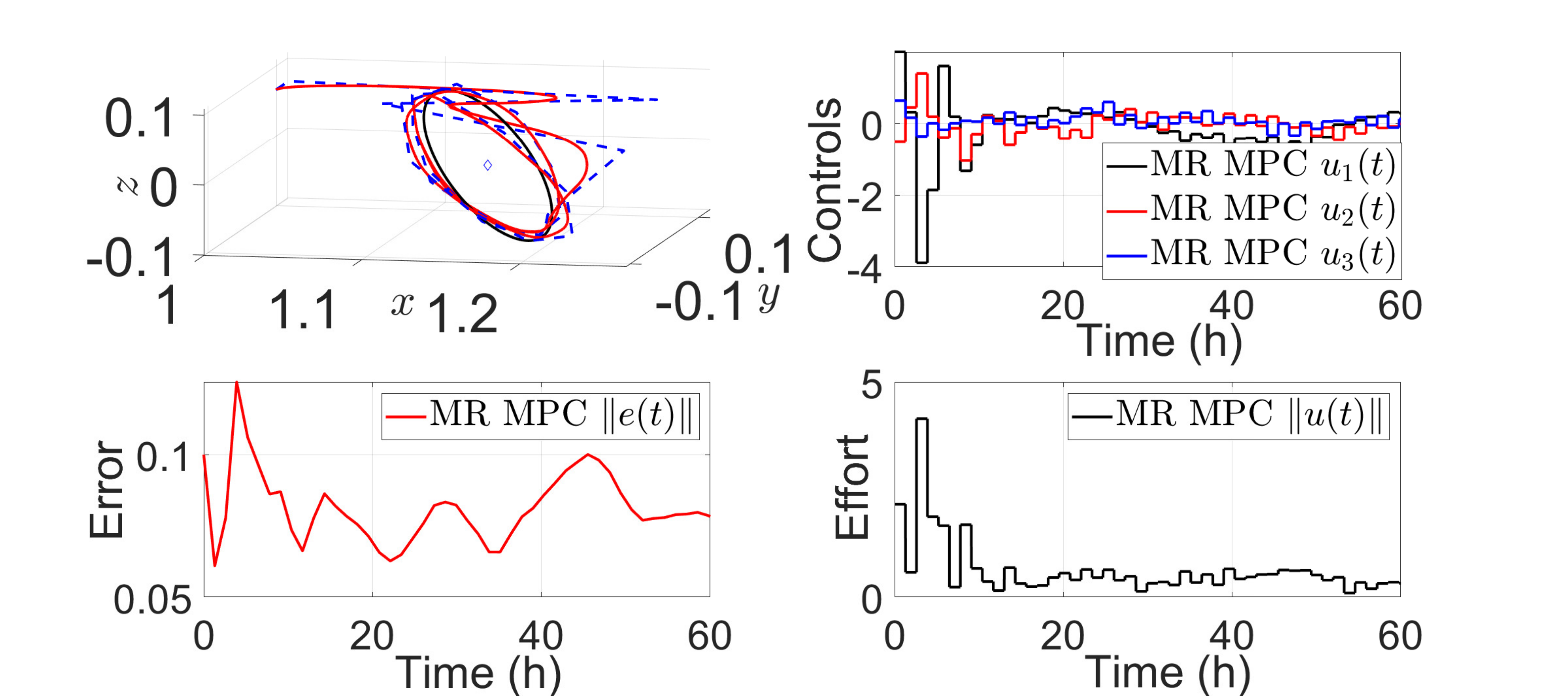}
\caption{Station-keeping under MR MPC with $\bar \delta = 1.2 \text{hr}$. All plots are in non-dimensional units.}
\label{sim_4MRMPC}
\end{figure}

A major benefit for the proposed MR MPC is the robustness with respect to higher sampling intervals. In fact Figure \ref{sim_4MRMPC} depicts how the proposed MR MPC maintains the spacecraft close to the quasi Halo orbit when increasing the sampling interval length to $\bar \delta = 1.2 \text{hr}$ in dimensional units (still lower than that reported in 4.A.7 in \cite{survey}), starting close to the orbit, namely $q(0) = (
1.022 \ 0 \ 0.12 \ 0 \ 0.1 \ 0
)^\top $ in non-dimensional units. Indeed, in this comparison we assume no control saturation. At this sampling rate, all other reported controllers fail to keep the spacecraft in the vicinity of the quasi Halo orbit.

\subsection{Computational aspects and RTI MPC}\label{scenario5}






To study the real time implementation question, we opt to utilize the real time iteration model predictive control \cite{Diehl}, together with the multirate sampled-data planner. The simulations are carried out using ACADO'S \cite{acado} Matlab Interface. Simulations were run on a PC running Windows 10 with an Intel i7 9$^{th}$ generation CPU and 16 GB of RAM. For testing reasons, Horizon lengths up to 40 steps were used for the 3 different simulation scenarios discussed in the previous section. 
Table \ref{table_4} shows the underlying Quadratic-Program (QP) solution time for a prediction horizon length of $n_p = 30$. This indicates that the MR MPC, with an RTI MPC solver can be implemented to solve the problem of station-keeping in real time provided some trade off is admissible in terms of quasi Halo orbit station-keeping performance.

\begin{table}[ht]
  \begin{varwidth}[b]{0.5\linewidth}
    \centering
    \begin{tabular}{ l r r }
      \toprule
      &Average time to QP solution in $\mu s$ \\ [0.4ex] 
      \midrule
Scenario 1 & 187      \\
 &     \\
 Scenario 2 &  191     \\
 &      \\
 Scenario 3 & 220     \\
      \bottomrule
    \end{tabular} 
    \caption{Time required by the CPU to solve the QP in the RTI in Scenario 1}
    \label{table_4}
  \end{varwidth}%
  \hfill
  \begin{minipage}[b]{0.5\linewidth}
    \centering
    \includegraphics[width=\textwidth]{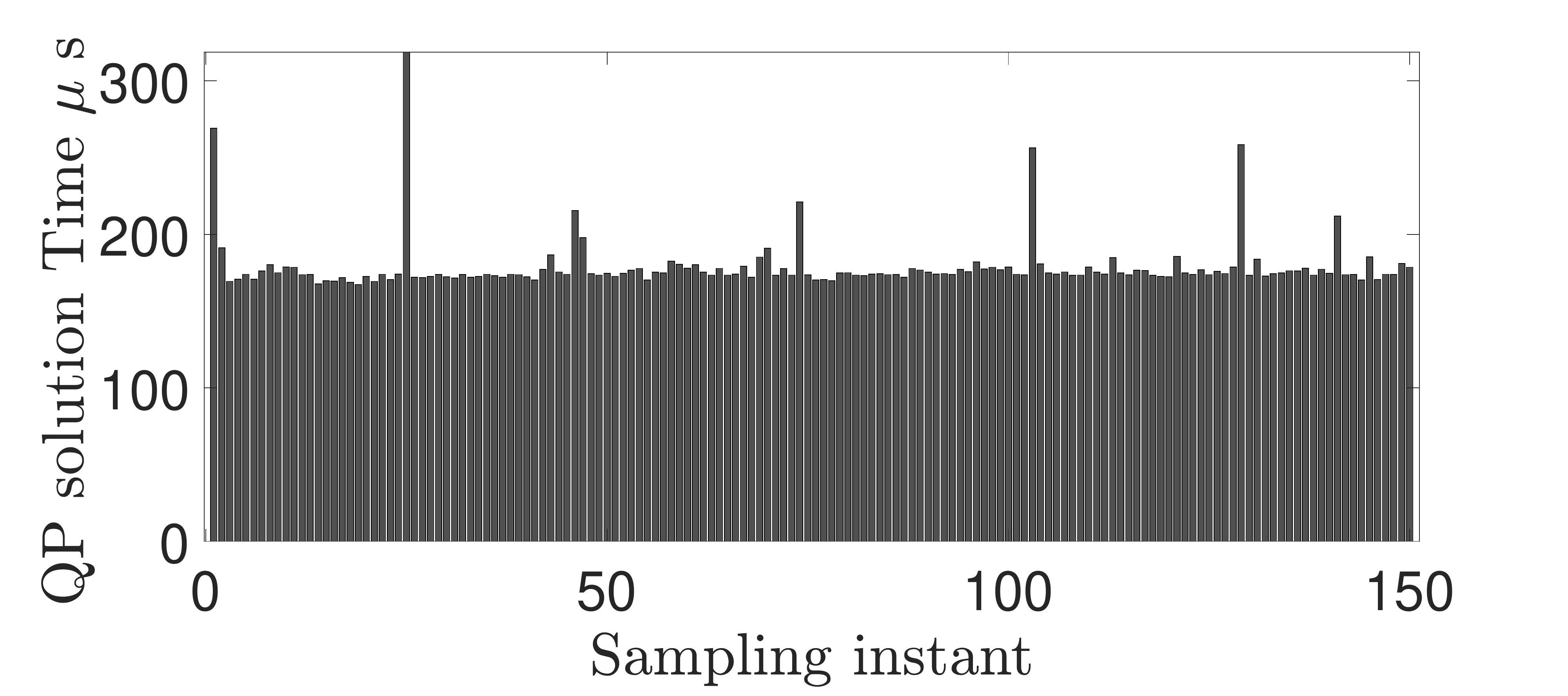}
  \end{minipage}
\end{table}

\section{Conclusions}\label{SEC:VI}

A sampled-data controller based on multi-rate planning and nonlinear model predictive control has been proposed for allowing a spacecraft to track a quasi Halo orbit around the $L_2$ point in the Earth-Moon system. In particular, the control is composed of two main components: a multi-rate trajectory planner which, based on a simplified feedback equivalent discrete-time model, generates admissible and bounded references guiding the spacecraft to the orbit; a model predictive controller which, based on the aforementioned trajectories and a suitably defined sampled-data prediction model, solves at each sampling instants a constrained optimization problem. 
Because of the generated references, such a problem is feasible and the closed-loop trajectories are bounded.

Simulations based on realistic scenarios highlight that the performances achieved under the proposed control scheme are significantly improved when compared to standard digital implementations (that is through zero-order-hold) of standard nonlinear control laws (e.g., based on regulation, feedback linearization control) and polynomial optimization based nonlinear MPC in station-keeping of quasi Halo orbits applications. 
Several aspects were considered such as eccentricity related perturbations and saturation limits on the control inputs. Key performance indicators enforce the validity of the proposed control scheme, both in terms of orbit tracking root-mean-square error and the $\mathcal{L}_2$ norm of the accelerations required for station-keeping as a measure of energy expenditure. Furthermore, robustness to both solar radiation pressure, as modelled by an additive periodic perturbation, and slower sampling rates is illustrated.  Finally, it results that the proposed scheme is suitable for real time digital implementation on modern hardware, even with longer prediction horizons for the MPC than the maximum typically assumed allowable in the literature (e.g. longer than $40$ time steps), through small adjustment to the online optimization solver.

Since the methodology here proposed is general in nature, an additional conclusion that can be drawn is that it can be used, with the appropriate changes, to handle a variety of Halo and Lyapunov orbits around libration points both in the Earth-Moon and Sun-Earth systems. Of course the results here reported are preliminary in nature albeit paving the way for sampled-data approaches in station-keeping applications.

\section*{Acknowledgments}

Mohamed Elobaid wishes to thank  Universit\'e Franco-Italienne/Universit\`a Italo-Francese (Vinci Grant 2019, Chapter II) for supporting his mobility between France and Italy within his PhD.  
The Authors wish to thank the anonymous Reviewers for their comments and suggestions which notably helped them in improving this work.
\bibliography{sample}

\end{document}